# DECISION MAKING IN ECONOMICS- A BEHAVIORAL APPROACH

## AMITESH SAHA

A project report submitted

In partial fulfillment of the requirement for the award of the degree of

**MASTER OF SCIENCE
IN
ECONOMICS**

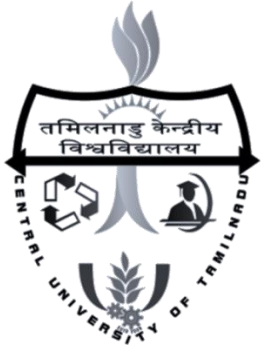
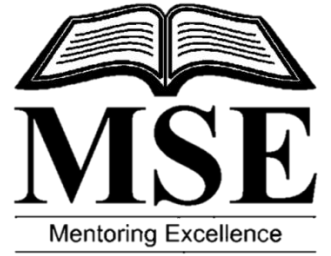

MADRAS SCHOOL OF ECONOMICS

AND

CENTRAL UNIVERSITY OF TAMIL NADU

APRIL 2017

MADRAS SCHOOL OF ECONOMICS
Chennai -600025

| | | |
|---|---|---|
| *Degree and Branch* | : | MASTER OF SCIENCE (GENERAL ECONOMICS) |
| *Month and Year of Submission* | : | APRIL 2017 |
| *Title of the Project Work* | : | DECISION MAKING IN ECONOMICS-A BEHAVIORAL APPROACH |
| *Name of the Student* | : | Amitesh Saha |
| *Roll Number* | : | MSE Roll No. 2015GE08 |
| *Name and Designation of Supervisor* | : | Dr. R. Hema<br>Associate Professor,<br>Madras School of Economics<br>Chennai - 600025 |

# BONAFIDE CERTIFICATE

Certified that this Project Report titled **Decision Making in Economics-A Behavioral Approach** is the bona fide work of **Mr. Amitesh Saha** who carried out the project under my supervision. Certified further, that to the best of my knowledge the work reported herein does not form part of any other project report on the basis of which a degree or award was conferred on an earlier occasion on this or any other candidate.

**Dr. Sunder Ramaswamy**                                                                  **Dr. R. Hema**
Director,                                                                                      Associate Professor,
Madras School of Economics                              Madras School of Economics
Chennai-600025                                                                                           Chennai-600025

# Acknowledgement

This work owes its existence to the guidance, motivation and support of many people. Firstly, I wish to express my sincere gratitude to my dissertation advisor Prof. R. Hema who had continually guided me throughout the work. Her support, advice and suggestions have been the most helpful in the writing of the dissertation. I am grateful to Dr. Venkatachalam L. for introducing me the idea of Behavioral Economics. I am also in debt to Dr. Santosh Kumar Sahu for his help and guidance, if it were not for the discussions I have had with them, this dissertation would not have had its current form. This would not have been successful if it were not for my friends' and my fellow students' constant motivation - the help of Varsha Choraria and Ayush Kumar was especially vital in completing the work. Finally, I must express my deepest gratitude to my parents for providing me with unfailing support and continuous encouragement throughout my years of study and through the process of getting through with education. This accomplishment would not have been possible without them.

Amitesh Saha

# Contents

# Abstract

We review economic research regarding the decision making processes of individuals in economics, with a particular focus on papers which tried analyzing factors that affect decision making with the evolution of history of economic thought. The factors that are discussed here are psychological, emotional, cognitive system and social norms. Apart from analyzing these factors it deals with the reasons behind the limitations of rational decision making theory in individual decision making and the need for a behavioral theory of decision making. In this regard, it has also reviewed the role of situated learning in decision making process.

# Introduction

Decision making is a process of identifying alternative courses of action and selecting an appropriate alternative in a given situation. This definition has two important parts: one, identifying alternative courses means that an ideal solution may not exist or might not be identifiable; two, selecting an alternative means that there may be a number of inappropriate alternatives that are to be rejected[1]. Thus, individual judgment on situations takes an important role in decision making. The idea of theorizing decision making has been a core problem to economists as well as psychologists since the evolution of thought.

To trace back to the history of economic thought we could start with Fan Li (517BC), an ancient Chinese advisor in the state of Yue, who first wrote on economic issues and developed 'golden' rules for business transaction. Following Fan Li, Chanakya (350BC) of India wrote Arthashastra on economic policies and military strategies[2]. In Greco-Roman World Aristotle criticized Plato (380BC-360BC) in his theory of money, explaining the possible 'wickedness of human nature'. The 'wickedness' later has been theorized as irrational behavior. During 500AD-1500AD there were Thomas Aquinas, Duns Scotus, Jean Buridan, Ibn Khaldun and Nicole Oresme. All of them mostly focused on 'just price' theories and cost of productions[3]. Antonin of Florance (1389-1459) was the first to think about character of a good and its value, based on how it satisfies human needs, instead of its nominal price. In 1662, Sir William Petty introduced statistical mathematics and gave a new dimension to the preceding tradition. In pre classical era, Richard Cantillon (1680-1734) in his essay on the 'nature of commerce in general' argued "rational self interest in freely-adjusting markets would lead to order and mutually compatible prices"[4]. Following David Hume, Bernard Mandeville (1670-1733) showed that the outcomes of the actions of men could not be separated into higher and lower forms based on worthiness. This idea was introduced only to ease the role of government and the relations of society. He defined 'virtue' as "every performance by which man, contrary to the impulse of nature, should endeavor the benefit of others, or the conquest of his own passions, out of a rational ambition of being good". In 1776, 'An Inquiry into the Nature and Causes of The Wealth of Nations' made Adam Smith the father of modern political economics. It not only explained the crux of American Revolution but also showed the path to new industrial revolution that oiled in creating more wealth in large scale[5]. The first book of Adam Smith, 'The Theory of Moral Sentiments' (1759) explained how individuals progressed their senses, learning from other individuals to the facts which can be termed as right and which can be termed as wrong. One of his famous statement on self interest was

---

[1] This definition is taken from 'Introduction to financial economics: A user perspective' by Jones, Warner, Terrell, Irvine, Allwright. It focused on improving student's decision making skills for their career and for better use of accounting information.

[2] Robin R. Wang in his book 'Yinang: The way of Heaven and Earth in Chinese Thought and Culture' mentioned about Fan Li who was the first person who wrote twelve golden rules for business transactions.

[3] 'Just Price' is a theory of ethics which tries to set fairness in transactions. The idea initiated in ancient Greek Philosophy and later advanced by Thomas Aquinas.

[4] This idea of Cantillon came from imitating Newton's force of inertia and gravity in the natural World with human reason and market completion in economic World during Britain's most troubling period of 17th century.

[5] Laura L. Frader mentioned a wide variety of primary sources in his publication 'The Industrial Revolution' including the role of Adam Smith.

> *"It is not from the benevolence of the butcher, the brewer or the barker, that we expect our dinner, but from their regard to their own self interest. We address ourselves, not to their humanity but to their self-love, and never talk to them of our own necessities but of their advantages."*

After Smith, Edmund Burke (1729-1797) wrote his book on 'Thoughts and Details on Scarcity' and Smith expressed an affinity to the opinion of Burke saying Burke was the only man whose thought on economic subjects were exactly same as Smith's. After Burke, Jeremy Bentham introduced the concept of utilitarianism to design a methodology for the calculation of aggregate happiness in a society[6]. 18th and 19th century is known as classical period and the first era when economics as a subject started being united. The ideas continued to grow with David Ricardo, John Stuart Mill, Karl Marx etc. In 1870, Neoclassical economics developed with three main independent schools of thoughts. The Cambridge School started developing theories on partial equilibrium and focusing on market failures. The Austrian school of economics developed theory of capital and tried to explain economic crisis and The Lausanne school developed the theories of general equilibrium and Pareto efficiency. William Stanley Jevons (18835-1882) in 1871 published 'Theory of Political Economy' stating that at the margin the satisfaction of goods and services decreases. The idea of diminishing marginal utility was seeded. With the evolution of thought, various thinkers from various schools started giving further insights on various dimensions. Friedrich Hayek (1899-1992) of the Austrian school disparaged the idea of 'social justice'. Thorstein Veblen (1857-1929) was one of the critics of consumerism and profiteering[7]. All these schools of thought were emerging in the backdrop of world wars, revolutions, depressions ( early to mid 20th century) and so on with their criticisms as well. There has been a thin demarcation line between classical and neoclassical economists. 'The Theory of Moral Sentiments' by Adam Smith proposed psychological explanations of human behavior, including concerns about fairness and justice. Jeremy Bentham also wrote psychological underpinnings of utility. However, the development of neoclassical economics reshaped the discipline as a natural science, *deducing* economic behavior from *assumptions* about the nature of economic agents. The emphasis was on the 'rational' character of human beings.

Throughout the evolution of economic ideas, 'theorizing human behavior' has been a crucial matter to economists. Economics has always been a study of choices under scarcity and uncertainty. Therefore, economists tried to hold on with assumptions and axioms that suit best in defining the choice behavior of sectors, where individuals are involved. Till the beginning of 20th century most of the neoclassical economists agreed to the traditional theory of rational human being in decision making theories. A rational human being is defined as one who always maximizes his satisfactions under any circumstances. There are three broader assumptions about the objectives of economic agents: one, rational consumers maximize their satisfaction or utility from consumption, by correctly choosing how to spend their limited income; two, producers/ firms wish to maximize profits, by producing at lowest cost the goods and services that are desired by consumers and three, government wishes to improve the economic and social welfare of citizens. Rational decision making model assumes that the best decision maker is rational and consistent while making value-maximizing choices with the following assumptions: a) able to define the problem, b) identify the decision making criteria, c) allocating weights to criteria, d) developing the alternatives (assumes we have all the information), e) evaluating the alternatives and f) selecting the best alternative. Undoubtedly, the rational decision making model is positive in nature but the responses of individuals do change under the influence of circumstances.

---

[6] In 'The principle of Morals and legislations (1791) with the aim of decreasing misery and sufferings Jeremy Bentham desired to focus on happiness.

[7] In 'The Huchinson Unarbridged Encyclopedia with atlas and Weather guide, Abington' it was mentioned that Veblen argued that there exists a class struggling under capitalism between businessman and engineers. his work prompted that life in modern industrial community is the result of a polar conflict between making money and making goods.

Herbert. A. Simon was the first economist to mention that most people are only partly rational and are seemingly irrational in the remaining part of their actions in 'Models of Man'[8]. He introduced the idea of bounded rationality to find more realistic mapping of human nature to their actions and challenged the assumptions of rational decision making theory. Simon pointed out that most consumers and businesses do not have sufficient knowledge and computational abilities to make fully-informed judgments while making their decisions. Because of limited information capacity, it is impossible to understand all the information. Bounded rationality suggests that consumers and businesses should focus on satisfaction rather than maximization. It's better to use simple rules of thumb to make decisions as the assumptions of rational decision making do not really hold in real day-to-day life. There are various factors that affect human beings while they make decisions. Behavioral economics studies the effects of these factors on the economic decisions of individuals. There is a vast literature which deals with individual factors that affect decision making. This paper reviews the economic research of such papers and tries to provide a brief idea about how psychological, emotional, cognitive and social factors affect the process of decision making in economics and whether there is any role for real-time learning in improving decision making. Section 1 gives a brief introduction about the origin of Simon's idea of bounded rationality. Section 2 considers research on psychological factors that affect decision making. Section 3 considers emotional effects in individual decision making. Section 4 reviews the cognitive system and the contribution of neuroeconomics in decision making. Section 5 considers the social norms that affect human actions and their decisions. Section 6 deals with the role education. Section 7 concludes.

---

[8]The book 'Models of Man: Social and Rational' by Simon combines multiple perspectives, primarily philosophical, economic and psychological, to create a model for rational human behavior in social setting.

# 1. The Starting Point

Herbert. A. Simon provided a critical assessment of the developments in economic science in the neoclassical rationalist tradition. He identified the shortcomings in this approach and presented an alternative framework for decision theory, based on a behavioral approach, which he and a few others had developed. He outlined the context which set him on a search for an alternative theory of decision making and how the behavioral foundations of decision theory evolved out of this. As Simon explained, the neoclassical approach assumed perfect rationality for each individual. The theory is amazingly simple and mathematically elegant but its predictive power for outcomes stems from controlling the environment. It assumes omniscient rationality on the part of individuals and that markets are characterized by perfect competition. Such controlled environment fully determines the behavior of individuals and organizations. The outcomes are based on equating marginal costs and marginal benefits. However, in reality we do not witness such perfect rationality nor do we find people actually making decisions based on marginalist principles. The behavioral approach to decision making assumes 'bounded' or limited rationality on the part of the individuals. These theories are based on modest and realistic demands on the knowledge and computational abilities of the human agents and they also fail to predict that human agents equate costs and returns at the margin.

## 1.1 A case in search of Descriptive theory

In 1934-35 Simon was engaged in a field study that looked at public recreational facilities in Milwaukee. These facilities were managed by the school board and the city public works department. While both these organizations were not competing for empire they were always disagreeing about the allocation of funds for physical maintenance and for play area supervision. They could not equate expenditures at the margin as neoclassical economics would recommend because they actually couldn't in reality. It was not possible to have a measurable quantitative production function from which marginal productivities could be derived. Also, the quantitative notions of the production function that each group had were mutually incompatible. In this case what kind of a decision rule could be adopted? Such a situation is common in human decision making. To tackle this, organizational theory resorts to 'sub-goal identification'. When goals cannot be operationally connected with actions, then decisions can be made on the basis of sub-goals that could be connected with actions. However, there is no unique way of formulating these goals. The formulation would depend on the knowledge, experience, the organizational environment as also maybe the self-interest and the power drives of the decision maker. The neoclassical approach in this context calls for complete knowledge of all possible alternatives and the ability to assess the consequences that would follow each of those. It expects certainty on the decision maker's current and future evaluation of these consequences. It requires the ability to compare these consequences using a consistent measure. Thus in a context where knowledge of the potential alternatives is incomplete and of the consequences of choosing particular alternatives is imperfect there is a need for an alternative decision rule. One procedure to tackle this is to look for 'satisfactory' choices rather than 'optimal' ones. Another would be to replace global goals with tangible sub-goals whose achievement can be observed and measured. A third approach would be to divide up the decision-making task among many specialists and coordinate their work through a structure of communications and authority relations. All of these would come under the rubric of bounded rationality. Simon hence perceived the elaborate organizations that human beings have constructed to carry out the work of production and government, as machinery for coping with the limits of human ability to comprehend and compute in the face of complexity and uncertainty.

## 1.2. Validation

Simon went into the question of whether there are empirically verified aggregate predictions that follow from the theory of perfect rationality but that do not follow from behavioral theories of bounded rationality. He takes up the cases of (i) negatively sloped demand curves, (ii) first degree homogeneity of production functions, (iii) shape of the long run cost curve and (iv) executive salaries.

### 1.2.1. Negatively sloped demand curves

There is no evidence on the ground that consumers actually equate marginal utilities with a view to maximizing their total utilities so as to determine how their income is distributed across their purchases. What the empirical data confirm is that demand curves are negatively sloped. However, having perfect rationality and a maximization objective are not the only behavioral conditions under which this outcome is possible. As Gary Becker (1962) has argued, the negative slope of demand curves largely results from the change in income opportunities alone and is largely independent of the decision rule. Thus, the utility maximization decision rule might be a sufficient condition but is not a necessary condition.

### 1.2.2. First degree homogeneity of Production Functions

Here again the neoclassical rationality assumptions provide sufficient but not necessary conditions. Fitted Cobb-Douglas production functions are homogeneous and have been found to be of degree close to unity and with the labor exponent to be of right magnitude. Such identical results can also be obtained by mistakenly fitting a Cobb-Douglas function to data that were to be generated by a linear accounting identity. The same applies to the Constant Elasticity of Substitution (CES) production functions. Behavioral theories of decision making would also be equally compatible with such data.

### 1.2.3. The long run cost curve

Neoclassical theory requires the long run average cost curves to be U-shaped if competitive equilibrium is to be stable. Theories of bounded rationality do not predict this and observed data on the ground indicate that costs at the high scale ends of the curves are either constant or even declining. These findings are compatible with stochastic models of firm growth and size rather than the static equilibrium model of neoclassical theory.

### 1.2.4. Executive Salaries

The average salaries of top corporate executives are found to increase proportionately with the logarithm of the size of the corporate. This finding can be arrived at based on the neoclassical model of profit maximization only when some ad hoc assumptions are made regarding the distribution of managerial ability. However, behavioral theory can arrive at this conclusion by assuming a culturally determined factor that fixes managers' salaries at a proportionately higher level compared to those of their immediate subordinates.

In all these cases, behavioral theory could predict aggregate outcomes as well as the neoclassical theory based on more realistic and plausible assumptions.

Simon states that phenomena that can be explained with the theory of utility or profit maximization but cannot be explained by the theory of bounded rationality have not been observed. However, there are situations that behavioral theory is better equipped to handle than the neoclassical theory and these involve decision making under uncertainty and imperfect competition. In these cases, neoclassical theory has provided enormous conceptual clarification by using statistical decision theory and game theory. However, these have not been able to describe human behavior under these conditions satisfactorily nor have they provided prescriptions that are usable given the limited computational powers of men and machines.

### 1.3. Another perspective

The comparison of behavioral theory with neoclassical theory so far has been framed in the context of building the foundations for political economy or for explaining phenomena that are intrinsically interesting. We now move on to looking at decision theory from the perspective of offering direct advice to businesses and governments. Decision tools employed by managers must enable them to actually make decisions based on the data and the limited computational capacity available at their disposal. In this context, there are two options: either the optimization techniques based on perfect rationality be retained but the environment is simplified so that the optimum is computable. Alternatively, satisficing models can be used that retain a richer set of the properties of the real world and provide good enough decisions. Hence, descriptive decision theory is about ways in which decisions are made rather than just about decision outcomes.

Why are most of the economic activities taking place in business firms that are characterized by employment relations within a hierarchical authority? Simon suggests that this is because the employer has uncertainty as to which future behaviors would be advantageous to him, while the employee has greater indifference as to which of these behaviors he carried out, within certain bounds. Thus under conditions of uncertainty it would be better to hold resources in 'flexible form'. Simon further explains organizational equilibrium in terms of motivational theory as the balance between inducements that are provided by the organizations to their participants and the contributions that these participants make to the organizations' resources. In principle the inducements-contributions argument is similar to the classical theory of the firm except that while the classical theory assumes that all profit will go to the owners the organizational theory treats the surplus more flexibly which leaves room for bargaining among the participants for a share in the surplus. The survival condition for the firm would be positive profits rather than the maximum profits that is required under perfect rationality.

Bounded rationality is basically considered as a residual, because if rationality falls short of omniscience then it is bounded. Two concepts are central to this characterization – 'search' and 'satisficing'. If the alternatives available to the decision maker are all not known to him then there must be a search for them. Hence a theory of bounded rationality must incorporate a theory of search. Also, utility maximization is not essential for a search scheme. Instead, the decision maker can have a certain aspiration about how good an alternative he must find. As soon as an alternative that meets his aspiration is found the search can be terminated. This notion of the aspiration level has its roots in psychological theory. These aspiration levels are not static and could vary with changing experiences. Aspirations rise in a favorable environment and fall in harsher environments.

## 2. Psychology and decision making

We now turn our focus to psychological factors that affect decision making. There is a large literature that deals with the possible reasons of how and why psychology of individuals clouds their judgments. Amos Tversky, Daniel Kahneman, Richard Thaler and Mathew Rabin are few of them who explained in details why human judgments are sometimes different from the way they are traditionally described by economists. Mathew Rabin suggested various modifications to the concept of human choices of stable preferences, that given a set of options and probabilistic beliefs, a person is assumed to maximize the expected value of utility function. A person's preferences are mostly determined by changes in outcomes relative to her reference level and not by absolute level of outcomes. Individuals are adaptive in nature. The outcomes that are best for a circumstance may totally be insignificant in a different circumstance. Therefore, Rabin suggested to focus on outcomes that are relative to reference levels. Both Mathew Rabin and Daniel Kahneman reasoned uncertainty behind the biases of individual's judgments. Individual judgments are mostly based on two systems; reasoning and intuition. Frederick (2003) has used a simple puzzle to monitor this two systems as in the following example: " A bat and a ball cost $1.10 in total. The bat costs $1 more than the ball. How much does the ball cost?" Almost everyone reports an initial tendency to answer "10 cents" because the sum $1.10 separates naturally into $1 and 10 cents, and 10 cents is about the right magnitude. Frederick found that many intelligent people yield to this immediate impulse: 50% (47/93) of a group of Princeton students and 56% (164/293) of students at the University of Michigan gave the wrong answer. Clearly, this responses were given without checking it. The surprisingly high rate of errors illustrated "people are not accustomed to thinking hard, and are often content to trust a plausible judgment that quickly comes to mind". Reasoning is what we do when we compute the product of 19 by 257 or fill an income tax form. Intuition is at work when we read the sentence "Bill Clinton is a shy man"[9].

## 2.1. The Accessibility in Behavior

Intuitive thoughts come to our mind spontsaneously. The technical term, for the ease with which mental contents come to our mind, is accessibility. To understand intuition, we must understand why some thoughts are accessible and others are not. Accessibility has many dimensions. At one end we find rapid, automatic and effortless operations and on the other end slow, serial and effortful operations that people need a special reason to undertake. Some of the determinants of accessibility are probably generic; other develop through experiences. The impressions that become accessible in any particular situation are mainly determined by the actual properties of the object to be judged. Physical salience also determines accessibility. If a large green letter and a small blue letter is shown at the same time, "green" will come first to our mind. However salience can be overcome by deliberate attention: an instruction to look for small objects will enhance the accessibility to all its features. The statements " Team A beat team B" and "Team B lost to team A" convey the same information, but because each sentence draws attention to its grammatical subject, they make different thoughts accessible.

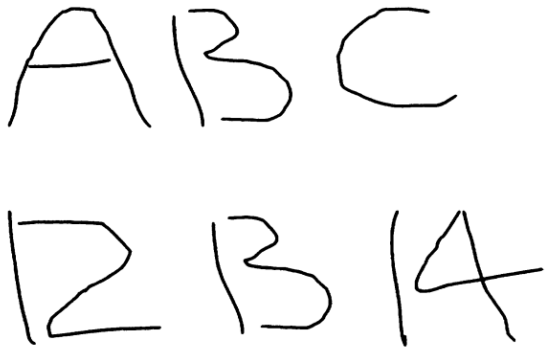

FIGURE 1. AN EFFECT OF CONTEXT ON ACCESSIBILITY

Figure 1 (adapted from Jerome S. Bruner and A. Leigh Minturn, 1995) includes a standard demonstration of the effect of context on accessibility. An inherent stimulus that is perceived as a letter within a context of letters is instead seen as a number when placed within a context of numbers. Expectations (conscious or unconscious) are a powerful determinant of accessibility. Proximity plays a vital role in this case. For the reader who sees the two versions in close

---

[9] Paul Rozin and CArol Nemroff, (2002), Daniel T. Gilbert, (1989), (2002); Timothy D. Wilson, (2002); Seymour Epstein, (2003)

proximity, the aspect of the demonstration is spoiled but when the two lines are shown separately, observers will not become aware of the alternative interpretation. The compound cognitive system, that has become sketched here, is an impressive computational device. However, this marvelous creation differs in important respect from another ideal, the rational agent assumed in economic theory. Some of these differences are explored in the following sections.

## 2.2. The Prospect Theory

Perceptions are dependent on references. This section explores the literatures of Tversky and Kanheman to show that intuitive evaluations of outcomes are reference dependent. Putting the hand in water at 20ºC will feel warm after keeping it in colder water, and will feel pleasantly cool after putting it in much warmer water. The point of demonstration is that, the experience of temperature is not a single parameter function of the temperature to which one is currently exposed. From the perspective of a student of perception, it is quite surprising that in standard economic analysis the utility of decision outcomes is assumed to be determined entirely by the final state of endowment, and is therefore reference-independent. Tversky and Kahneman constructed a large number of experiments that led them to the formulation of prospect theory ( Kahneman and Tversky, 1979). One ornament of prospect theory is that, it was explicitly presented a formal descriptive theory of the choices that people actually make and not as normative model. This was a departure from a long history of choice models that served a great deal as logics and idealized descriptive models.

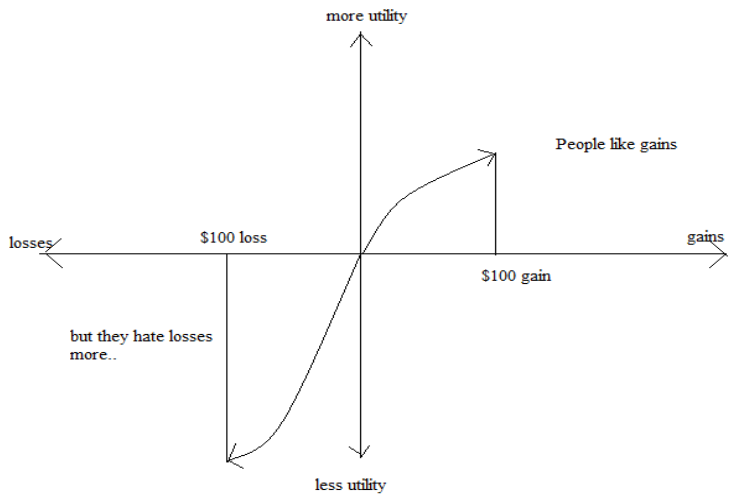

FIGURE 2. A VALUE FUNCTION

The prediction of prospect theory follows from the shape of value function as shown in figure 2. The value function is defined on gains and losses and is characterized by three features: a) It is concave in the domain of gains, favoring risk aversion; b) It is convex in the domain of losses, favoring risk seeking; c) The function is sharply kinked at the reference point, and loss-aversion is steeper for losses than for preferences by a factor of about 2-2.5.

## 2.3. Framing Effects

In the context of decision making, the assumption that preferences are not affected by inconsequential variations in the description of outcomes, is considered as an essential aspect of rationality. Invariance is violated in framing effects, where descriptions lead to different choices by altering the relative salience of different aspects of the problem. Tversky and Kahneman (1981) introduced their discussion of framing effects with the following problem:

"Imagine that the United States is preparing for the outbreak of an unusual Asian disease, which is expected to affect 600 people. Two alternative programs to combat the disease have been proposed. Assume that the exact scientific estimates of the consequences of the programs are as follows: If the program A is adopted, 200 people will be saved. If program B is adopted, there is a one third probability that 600 people will be saved and a two thirds probability that no people will be saved."

In this version of the problem, a large number of respondent favored program A, indicating risk aversion. Other respondents, selected at random, received a question in which the same cover story is followed by a different description of the options:

"If program A is adopted, 400 people will die. If program B is adopted, there is one third probability that nobody will die and a two thirds probability that 600 will die."

Now majority of respondents favored program B, the risk seeking option. Although there is no substantive difference between the two problems, people responded and evaluated differently. Thus it can be seen how in certain options, outcomes that are certain, are over weighted, relative to the outcomes of high or intermediate probability. Thus, the certainty of saving people is disproportionately attractive, while accepting the certain death of people is disproportionately aversive. There has been considerable interest among behavioral economists in a particular type of framing effect, where a choice between two options A and B is affected by designating either A or B as a default option. The default option is taken in such choices even when the alternative option has larger advantage. The basic impact of framing is the passive acceptance of the formulation given. Because of this passivity, invariance with respect to framing cannot be achieved by a finite mind. The potential of such invariance raises significant doubts about the descriptive realism of rational choice models.

# 3. Role of Emotions in decision making

We, humans are emotional creatures. Emotions sometimes play a crucial role in the errors we make while making decisions. Sometimes we know what is best decision to make but due to emotions (like anger, sorrows) we do take wrong decisions. There has not been much work on the influence of emotions in economics till 20th century. Jon Elster felt that, to understand the topic, one should focus on the specific issues than generalizing as a whole. Even in psychology until recently the focus was on cognitive behavior. There is a difference between emotion and behavior. Where economists focus on behavior, emotion theories try to explain emotions. The psychological approach has not focused on how emotions generate behavior. The main question that Elster tried to answer is how can emotions explain behavior?

To define the functioning of emotions in decision making, Elster categorized emotions in different ways. A. Cognitive antecedents: Emotions that are different from non rational factors, like pain and hunger etc. are triggered by beliefs. B. Intentional objects: A causes B to hurt C, in which case, C's anger may be directed at B or at A. C. Physiological arousal: Emotional states are characterized by hormonal changes and by changes in the autonomic nervous system. When we say we are afraid, we may refer to the aroused emotional state or simply to a complex of beliefs and desires, as when we say we're afraid it's going to rain ( Robert Gordon 1987) D. Physiological Expressions: Many expressions are directly functional, like smiling, blushing etc. and many others are byproducts of action patterns, like loudness of voice or compression of lips, for instance, do not enhance coping but follow from the general organism against danger. E. Valence:  This term is used to scale emotions in pleasure-pain axis. Emotions that are high on arousal (embarrassment) may be low in valence and vice-versa (boredom). E. Action tendencies: The action tendency of shame is to hide or disappear; that of guilt to confess etc. Action tendencies may lead to immediate actions or be modified by one of several regulatory systems (choices are involved).

Emotions sometimes are valuable (love) but sometimes are undesirable (shame). Hence having them or avoiding them could be the object of rational choice. Choosing emotional expressions are partially dependent on agents.

There are two ways of choosing emotions: one, occurrent emotions that occurs fully and emotional dispositions that do not occur fully. It occurs in steps and so that it takes time. The experience of shame for instance is unpleasant, so everyone tries to avoid it. one can develop disposition by analyzing the steps when it comes to inform it's affects to the people we know. Emotions can be chosen if there  are constraints on choice. For emotions like anger there may be "a point of no return". There can be seeking out of favorable occasions, like going out for a movie, that may have predictable experience based  on beliefs. Similarly we can avoid unfavorable occasions to avoid certain emotions. There is a tradition in philosophical and religious thoughts that wise men can control the occurrent or 'sudden expressions' of their emotions. They can shape their emotional dispositions. People go to psychotherapy because they are worried about their emotional reaction patterns. One can also manipulate other's emotion by shaping their beliefs.

Emotions improve decision making in two ways; 1. Rather than making optimal decision, it helps to express what matters. 2. Sometimes emotions can help us make best decisions. As mentioned by Elster, Ronald de Sousa (1987) said in many situations rational choice theory was indeterminate. It does not allow an optimal action. Sometimes, in emergencies, we cannot take mechanical decisions because our emotions cloud our cognitive thinking. An analysis of patients based on general neuropsychological data, Elster referred Damisio (1994) who drew conclusion that defective decision making capacity arose due to their lack of emotions.

The interaction between emotion and interest has led to cost benefit model of emotions. Guilt is itself a cost. Even if I don't have money with me, a beggar's visibility induce feeling of guilt. Guilt also induces costly behavior. If I had money, I would give some to the beggar to decrease the guilt. More accurately, I would give the amount of money where the marginal utility of money equals my guilt. Elster assesses the usefulness of this approach with respect to guilt, shame, envy, indignation, love etc. 'Dissonance theory' is more realistic than cost and benefit analysis. It views individuals making decisions based on reasoning instead of how they feel because people have an image about themselves that they are doing things for a reason.

## 4. An introduction to Neuroeconomics

Neuroeconomics tries to explain human decision making by understanding the functioning of the brain and its impact on human behavior. It studies how human behavior can shape our understanding. Hence, neuro-scientific discoveries can impact models of economic theories.

Simon analyzed and explained the reasons and consequences of the decision making process. To be more precise, he focused on intuition in human behaviors and how people think and make use of intuitions in decision making that involves inter personal interactions. According to Barnard, "logical process" is based on the ability of reasoning while "non-logical process" is influenced by judgments. In logical process, goals and alternatives are made explicit, the consequences of pursuing different alternatives are calculated, and these consequences are evaluated in terms of how closely they are to goals. In the judgmental decision making, the responses are rapid. Barnard also grounded the sources of these non logical processes. The direct sources are psychological, physical and social environmental factors. These are directly influenced by facts, patterns, concepts, techniques, abstractions and beliefs. Education, experiences and analyses also indirectly influence the judgments.

Simon went deeper to find the origin. He analyzed research on split brain[10] and found that the two hemispheres of the brain have qualitatively different functions. There is division of labor

---

[10] Two works that examine the split brain theory and forms of thought are R. H. Doktor's "Problem solving styles of executives and Management Scientistss" in A. Charnes, W. W. Cooper, and R. J. Niehaus's (eds.) Management Science Approaches to Manpower Planning and Organization

here : the right hemisphere, which controls the left side of the body, recognizes visual patterns and the left hemisphere which controls the right side of the body, plays role in analytical processes. Electrical activity in the brain can be measured by EEG (electroencephalogram) techniques. Activity in brain hemispheres is associated with partial or total suppression in the hemisphere of the alpha system, a salient brain wave with a frequency of 10 waves per second. When a hemisphere is inactive, the alpha rhythm in that hemisphere becomes strong. For most right-handed subjects, when the brain is involved in a task like say, recognizing visual patterns, the alpha rhythm is relatively strong in left than the right hemisphere and with more analytical tasks the alpha rhythm is relatively stronger in the right hemisphere.

Simon thought that chess experts, great physicists and chemists, etc. exhibit reasoning or analytical process combined with other processes for accessing knowledge banks. Similarly, an experienced manager has a large amount of knowledge in his/her memory. It is gained from training and experience and organized in terms of chunks and associated information. Marius J. Bowman has constructed a computer program capable of detecting company problems from an examination of account statements. Simon mentioned R. Bhaskar who gathered thinking-aloud protocols to analyze a business policy case[11]. From these and other research on problem solving and decision making by experts, we can draw two main conclusions: first, experts, often arrive at problem, diagnoses and solves rapidly and intuitively without being able to report how they attained the result; second, this ability is best explained by postulating a recognition and retrieval process, that employs a large number- generally tens of thousands -of chunks or patterns stored in long term memory. When the problem is more than trivial, the process has to be organized in a coherent way and they must be supplied with reasoning capabilities that allow inferences to be drawn from the information retrieved, and numerous chunks of information to be combined. Hence, intuition is not a process that operates independently of analysis; rather the two processes are complementary components of effective decision making process. We can improve judgments by specifying the knowledge and recognition capabilities, that expert in a domain need to acquire or by providing the human decision maker with an expert consultant or guide or sort.

What managers think they should do, either by analysis or by intuition, is very often different from what they actually do. A very common problem is postponement of difficult decisions. As Simon said "When people have to choose between two evils, they do not behave like Bayesian statisticians, weighing the bad against the worse in the light of their respective possibilities. Instead they avoid the decision, searching for alternatives that do not have negative outcomes". Often, uncertainty is the source of difficulty. Most of us could probably agree that 'blame avoiding' behavior is far more common than 'problem solving' behavior after a serious error has been made.

---

What all of these decision making situations have in common is stress, a powerful force that can divert behavior from the urgings of reason. The intuition of the emotion-driven manager is very different from that of the experts. The latter's behavior is the product of learning and experience, and is largely adaptive. The former's behavior is a response to more primitive urges, and is more often than not inappropriate. The response to any problem looks both forward and backward. It looks backward to establish responsibility for the difficulty and to diagnose it, and forward to find a course of action to deal with it.

Failure to give sufficient attention to the future stems from two causes. The first is interruption by current problems that have more proximate deadlines and hence seem more urgent; the second is the absence of sufficient "scanning" activity, that can pick up cues from the environment, that long term forces not impinging immediately on the organization have importance for it in the future. It's wrong to contrast "analytic" and "intuitive" styles of management. Intuition and judgment are simply analyses, frozen into habit and into the capacity for rapid response through recognition. Every manager needs to analyze problems systematically and to be able to respond to situations rapidly, a skill that requires the cultivation of intuition and judgment over many years of experience and training.

## 5. Social factors

Almost everyone lives in a society that has different rules and regulations with respect to geographical locations and beliefs accepted by most individuals living there. Social norms have an important role in individual behavior as well as in their decision making. To think what is appropriate and what is not, we often depend on expectations of behaviors by others. Social norms thus influence our preferences and behavior. How much to drink at a party, whether to join the yoga class or fitness class, how much to eat and so on are all decisions that are all partly guided by social norms. Various studies tried to find out the correlation through various kinds of meta analysis that, if there is any influence at all in our decision making of social norms or not. Jon Elster mentioned whether these norms are followed out of self-interest or not. When people obey norms, they have a particular outcome in mind. When norms are internalized, they are followed even when they are not really warranted. Elster wrote "..one does not pick his nose when he can be observed by people on a train passing by, even if he is confident that they are all perfect strangers whom he will never see again...".

We can categorize four aspects in this regard. (1) Social thoughts can be considered in two ways, a descriptive way and a prescriptive or injunctive way. Descriptive way specifies the behavior of others. It focuses more on their beliefs than their individual actions. Therefore, people who live in the society may follow such rules without giving it much thought. Prescriptive or injunctive way focuses on what is approved and provides more attention on attitudes of people associated with the group. We can expect that descriptive norms has larger effect on individual behavior but a smaller effect on attitudes than an injunctive norm (2) Concrete information of norms has a greater influence because it is more engaging and memorable than abstract information (3) If we have trust or a sense of identification, so that we can identify with the person or institute which provides these norms, the norms become more relevant and influential (4) Behavior in public can be noticed and corrected by others. Therefore, behavior in public is more influential than that in private.

## 6. The role of education

After the "cognitive psychology" revolution in 1960, psychologist paid a great deal of attention to education as an area of application. Simon provided claims of 'situated learning' identified in his National Research Council, 1994 report. The core idea of situated learning is that most of learning comes from specific situations. The emphasis should be given to the relationship between what is learned in the classroom and what is needed outside the classroom, and this has been a valuable contribution of situated learning. Simon proposed the following claims:

**CLAIM 1:** The potentials of any action cannot be fully described independently of the specific situation. The common examples are that kids can do calculations well when it comes to shopping, selling products in street and so on. However, they cannot do the same calculations in the classrooms. Simon mentioned Godden and Baddeley who found that divers had difficulty remembering under water what they learned on land or vice-versa. Sometimes knowledge is necessarily bound to a specific context by the nature of instruction.

**CLAIM 2:** The second claim is a corollary of the first. If knowledge is wholly tied to the context of its acquisition, it will not transfer to other context. However, even without assuming extreme contextual dependence, one could still claim that there is relatively little transfer beyond nearly identical tasks to different physical contexts. Simon pointed out the research of Weber (1884) and Fechner (1858) in this regard. They showed in research on 'transfer in psychology' that there could be either large amount of transfer, modest amount of transfer, small amount of transfer and no transfer at all or even negative transfer. How much there is and whether transfer is positive depends on the experimental situation and the relation of the material originally learned to the transfer material. Simon also mentioned Singley and Anderson (1989) who showed that transfer between tasks was a function of the degree to which the tasks share cognitive elements. A number of studies converge on a conclusion that transfer is enhanced when training involves multiple examples and encourage learners to reflect on the potential for transfer.

**CLAIM 3:** Training by abstraction is of little use. Abstract instruction can be ineffective if what is taught in the classrooms is not what is required in on the job. One needs to create better correspondence between job performance and abstract classroom instruction, and sometimes this means changing the nature of the job and fighting unwanted effects of apprenticeship learning. Abstract instruction leads to successful transfer, while concrete instruction can lead to failure of transfer. In a research of Biederman and Shiffrar (1987), Simon found that 20 minutes of abstract instruction brought novices up to levels of experts who had years of practice. Most modern information processing theories in "cognitive psychology" are "learning by doing" that implies that learning would occur best with a combination of abstract instruction and concrete illustrations of the lessons of this instructions. Abstract instruction combined with concrete examples can be a powerful method if learning is applied to a wide variety of future tasks.

**CLAIM 4:** Learning is inherently a social phenomenon and it should be done on complex problems. Let's consider the skills required to become a successful tax accountant. While an accountant must learn how to deal with clients, it's not necessary to know the use of a calculator. It's better to train independent tasks separately because fewer cognitive resources will then be required for performance, thereby resolving adequate capacity for learning. A large body in psychology shows that part training is often more effective when the part component is independent, or nearly so, of the larger task. A number of detrimental effects arise from cooperative learning- "free rider", "sucker", "status differential" and "ganging up" effects. The evidences show that skills in complex tasks, including those with larger social components, are usually best taught by a combination of training procedures involving both whole tasks and components and individual trainings and training in social settings.

The development from behaviorism to cognitivism was an awakening to the complexity of human cognition. Situated learning has raised consciousness to certain aspects of learning that were not fully appreciated in education. However, this consciousness-raising has had its negative aspects as well but misguided practices can have the appearance of a basic scientific research.

## 7. Conclusion

In this review, we tried to show the effects of factors that hinder our rational decision making. While rational theory assumes a favorable environment behavioral approach focuses on more realistic assumptions. Various experiments have suggested that individuals prefer lesser information to make choices. As the number of alternatives increases the difficulty in choosing the best alternative also increases. As Drew Fudenberg said ".. I believe that behavioral economics has many insights and observations that should be used to improve economics as a whole..".

Human brain is largely unexplained till date. There are many untold stories that have not been discovered yet. Greater insights into the human cognitive system will help us understand how judgments are formed and how choices are made based on them. Psychologists have also contributed a lot in this regard. The experiments and contributions of Kahneman and Tversky on decision making under uncertainty can be viewed as most influential. Their inspiration has played the vital role behind a booming research in behavioral economics and finance. They have had a substantial impact in other fields also. A current wave of research on the combined tradition of psychology and experimental economics is helping us gain more insights into such aspects as 'bounded rationality', 'limited self interest', 'imperfect self control' and so on which are important factors influencing a range of market outcomes. It is likely that parsimonious behavioral theories consistent with the evidences, may eventually replace the rational model of decision making in economics.